\documentclass{iaus}
\usepackage{graphicx}

\title[Isotopic Abundances of Carbon and Oxygen] %% give here short title %%
{Isotopic Abundances of Carbon and Oxygen in Oxygen-Rich
 Giant Stars} 

\author[T. Tsuji]   %% give here short author list %%
{Takashi Tsuji$^1$}

\affiliation{$^1$Institute of Astronomy, The University of Tokyo,
 Mitaka, Tokyo, 181-0015 Japan \break email:
ttsuji@ioa.s.u-tokyo.ac.jp}

\pubyear{2007}
\volume{239}  %% insert here IAU Symposium No.
\pagerange{1--3}
\date{?? and in revised form ??}
\jname{Convection in Astrophysics}
\editors{F. Kupka, I. W. Roxburgh \& K. L. Chan, eds.}
\begin{document}

\maketitle

\begin{abstract}
 $^{16}$O/$^{17}$O and $^{12}$C/$^{13}$C ratios in 23 M giants
are determined from high resolution IR spectra. The results 
are confronted with the current models on the convective mixing.

\end{abstract}

{\bf Introduction:} 
It has been known that CNO and their isotopic abundances are 
useful probes of the mixing in evolved stars.
 It is, however, by no means clear yet
if the observed results could be consistent with the predictions
of the stellar evolution models. Here, we concentrate on oxygen-rich
giants, possibly experienced the first and second dredges-up, but may not
be disturbed by the further processes such as the third dredge-up.
For this purpose, we extend our previous analyses on
CO spectra in M giant stars to including isotopic ratios, and we also 
hope to reexamine the carbon abundances. 

{\bf  The  Basic Stellar Parameters: }
We selected  23 red giant stars, for which 
the effective temperatures  are determined by the use of the
infrared flux method and the bolometric luminosities by the
integration of the SEDs. With the Hipparcos
parallaxes, absolute bolometric magnitudes are determined.
The error bars of $M_{\rm bol}$ are based on those of the parallaxes, 
but errors in  photometric data and in $T_{\rm eff}$ (about $\pm 0.1$\,mag
and $\pm$100\,K, respectively) are not included. 
Our sample is shown on the HR diagram  and compared with the
evolutionary tracks by \cite{Claret04} in Fig.\,1.
Unfortunately, nearly half of our sample cannot be accounted for by
the models of Claret, and we had to extrapolate the evolutionary tracks 
to  estimate stellar masses given in Table 1.

\begin{figure}[hb]
\begin{center}
\hspace{-5mm}
\includegraphics[height=50mm]{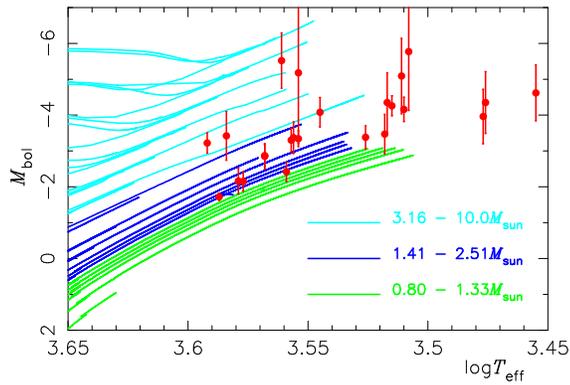}
\caption{Our sample on the HR diagram 
and the evolutionary tracks by Claret (2004).
}
\label{Fig1.eps}
\end{center}
\end{figure}

\begin{table} [hb]
\begin{center}
\caption{Isotopic Ratios and Basic
Stellar Parameters in 23 Red Giant Stars}
\begin{tabular}{ r c c  l l  c l c c c }
\hline \hline
\noalign{\smallskip}
\noalign{\smallskip}
   Obj.  &  $^{12}$C/$^{13}$C & $^{12}$C/$^{13}$C
 & $^{16}$O/$^{17}$O & $^{16}$O/$^{17}$O & log\,$A_{\rm C}$  & 
$\xi_{\rm mic}$ &  $T_{\rm eff}$ & $M_{\rm bol}$  &  $Mass$ \\ 
   &  present    & others   & present  & others &  & km\,$s^{-1}$ & 
K &  & $M_{\odot}$\\   
\noalign{\smallskip}
\hline
\noalign{\smallskip}
  RZ Ari  &  ~5 $\pm$  1 & &  ~402 $\pm$  57 & &  -4.09  & 
 2.99 &   3295 &  -3.47 $\pm$  0.55 &  1.5\\
 $\alpha$ Cet  &  10 $\pm$  2 & & ~200 &  &  -3.88  &  
 3.68  &     3905 & -3.22 $\pm$  0.28 &  3.6\\
  $\rho$ Per &     ~9 $\pm$  2 & 15 $\pm$  2$^a$ &  1000 & & -3.76 
  & 2.86 &   3505 & -4.08 $\pm$  0.41 &  3.2\\
 $\tau^4$ Eri &     ~9 $\pm$   2 & & ~560 $\pm$   97 & &
-3.67  & 2.29 &   3700 & -2.85 $\pm$   0.35 &  2.0 \\
  $\alpha$ Tau &  11 $\pm$ 2 & 10 $\pm$ 2$^a$ &   1000 & ~560 $\pm$
 180$^b$ & -3.71  & 2.59 &   3860 & -1.72 $\pm$  0.10 &  1.4  \\
\noalign{\smallskip}
  $\mu$ Gem &   ~7 $\pm$  1 & 13 $\pm$ 2$^a$ & ~206 $\pm$  30 & ~325
 $\pm$ 112$^b$ & -4.15  &  3.30 & 3605 & -3.30 $\pm$ 0.33 & 2.1\\
  $\nu$  Vir &  ~7 $\pm$ 1 & 12 $\pm$ 2$^a$ &  1500 & & -4.12 
 & 2.84   & 3795 & -2.16 $\pm$   0.36 &  1.6 \\
  $\delta$ Vir &   10 $\pm$  2 & 16 $\pm$ 4$^a$ & 1500 &  &-3.91
  & 3.33  &  3625 & -2.42 $\pm$ 0.27 &  1.4 \\
  SW Vir &   11 $\pm$   2 & & ~188 $\pm$ 22 & & -4.06  & 4.13
 &   2990 &  -4.35 $\pm$ 0.86 &  1.4 \\
  10 Dra &   11 $\pm$ 2 & 12 $\pm$ 3$^a$ &  ~112 $\pm$  14 &  & -3.77
  & 3.44 &   3700 &  -2.87 $\pm$  0.31 &  2.0 \\
\noalign{\smallskip}
  RX Boo &   ~5 $\pm$   1 & &  ~136 $\pm$  21 & &-3.75  & 
3.08 &   2850 & -4.62 $\pm$  0.78 &  1.4 \\
  RR UMi &    ~5 $\pm$  1 &  & 1000 & &  -4.10  & 
3.04 &   3355 &  -3.38 $\pm$  0.32 &  1.5 \\
 $\sigma$ Lib &    ~7 $\pm$  1 & &  1500 & & -4.00  & 3.21 & 
  3600 & -3.37 $\pm$ 0.44  & 2.1 \\
$\delta$ Oph &     ~9 $\pm$  1 & &  ~232 $\pm$  46 & & -3.97
 & 3.23 &   3775 & -2.15 $\pm$ 0.27 &  1.5 \\
 30g Her &    ~9 $\pm$   1 & 10 $\pm$ 2$^a$ & ~197 $\pm$ 24 & ~675 $\pm$ 
175$^c$ & -4.17  & 3.89 & 3235 & -4.16 $\pm$ 0.34 & 1.8 \\
\noalign{\smallskip}
 $\alpha$ Her &   ~9 $\pm$  1 &  & ~107 $\pm$  14 & ~190 $\pm$ 40$^b$ &
-3.93 &  3.81 & 3220 & -5.77 $\pm$  1.64 &  4.5 \\
  OP Her &    ~9 $\pm$  1 &  & ~184 $\pm$  21 & &  -4.12  & 
4.02 &   3285 &  -4.35 $\pm$  0.83 &  2.1 \\
  BS6861 &    21 $\pm$  5 & &  1000 & & -3.43  & 3.43 &  
 3580  & -5.18 $\pm$ 1.98 &  6.1 \\
 XY Lyr &    11 $\pm$ 1 &  &  ~218 $\pm$ 18  & & -4.10  & 
3.71 &   3245 & -5.09 $\pm$   1.05 &  3.4 \\
 $\delta^2$ Lyr &   11 $\pm$ 2 &  & ~199 $\pm$ 20 & & -4.21  & 
 4.51 &   3637 & -5.52 $\pm$  0.77 &  7.3 \\
\noalign{\smallskip}
  R Lyr &    ~8 $\pm$ 2 &  & ~376 $\pm$ 61 & & -3.73  & 
2.76 &   3275  & -4.26 $\pm$   0.28 &  2.0 \\
 $\lambda$ Aqr &    ~9 $\pm$ 1 & &  ~500 & & -3.87  & 
3.11 &   3835  & -3.42 $\pm$  0.68 &  3.6 \\
 $\beta$ Peg &  ~6 $\pm$ 1 & ~8 $\pm$ 2$^a$ &  1500 & 1050 $\pm$ 375$^b$
 & -4.11  & 3.39 &   3580 & -3.34 $\pm$  0.22 &  2.1 \\
\noalign{\smallskip}
\hline \hline
\noalign{\smallskip}
\end{tabular}
\vspace{2mm}
References: a) Smith \& Lambert (1990), b) Harris \& Lambert (1984),
c) Harris et al.(1985)  
\vspace{-2mm}
\end{center}
\end{table}

{\bf Method of Analysis:  }
We use the high resolution spectra observed with the KPNO FTS, and
measured EWs of isolated lines of $^{12}$C$^{16}$O, $^{13}$C$^{16}$O, and
$^{12}$C$^{17}$O in the $K$-band region. We applied a line-by-line 
analysis on these EWs with the use of the spherically extended 1D model
photospheres. In our analysis, the micro-turbulent velocity 
$\xi_{\rm mic}$ is determined so that the abundances of $^{12}$C$^{16}$O  
from individual lines show the least dependence  on the EWs.
The resulting $\xi_{\rm mic}$ is used to determine abundances of 
$^{13}$C$^{16}$O and $^{12}$C$^{17}$O as well.
Some examples are shown in Fig.2.

\begin{figure}[ht]
\begin{center}
\hspace{-5mm}
\includegraphics[height=45mm]{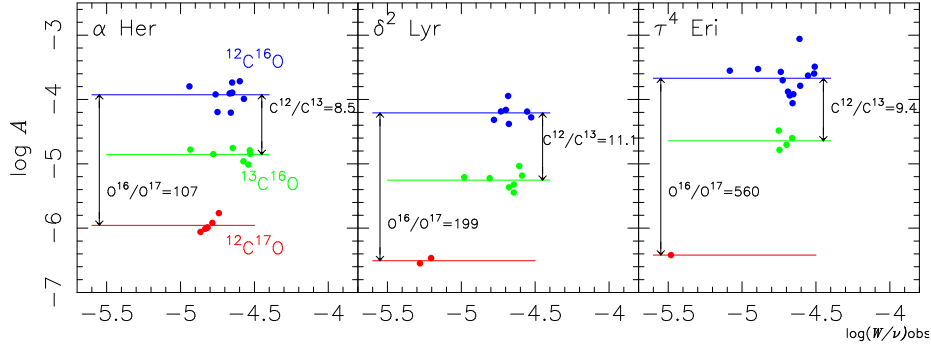}
\caption{ Line-by-line analysis on $^{12}$C$^{16}$O, 
$^{13}$C$^{16}$O, and $^{12}$C$^{17}$O in $\alpha$ Her,
$\delta^2$ Lyr, and $\tau^4$ Eri
}
\label{Fig2.eps}
\end{center}
\end{figure}

{\bf  Results:}  
The isotopic ratios based on the $^{12}$C$^{16}$O, 
$^{13}$C$^{16}$O, and $^{12}$C$^{17}$O abundances are given in Table 1. 
The results are further confirmed by a direct
comparison of the observed spectra with the synthetic spectra
in the regions of the $^{13}$C$^{16}$O (2-0) bandhead  
and $^{12}$C$^{17}$O (2-0) $R\,18 - 33$ lines. 
Also, we applied the spectral synthesis method to estimate 
$^{16}$O/$^{17}$O for the case in which $^{12}$C$^{17}$O lines 
are too weak to measure EWs accurately, and the results are given without
 error bars in Table 1. For a few stars, our results appear to be consistent
with the results by other authors (\cite[Harris \& Lambert 1984]{Harris84}, 
\cite[Harris, Lambert \& Smith 1985]{Harris85}, \cite[Smith \& Lambert 1990]
{smith90}) in general, as shown in Table 1.

{\bf $^{16}$O/$^{17}$O Ratio:}
The resulting  $^{16}$O/$^{17}$O ratios 
plotted against the stellar masses are compared
with the predicted ones by \cite{Claret04} (solid lines) in Fig.\,3a.
The rather large variation of $^{16}$O/$^{17}$O in low mass stars
is well consistent with the prediction of the evolutionary models, 
confirming  the previous analysis by \cite{Dearborn92} using the 
observed data known at that time. 
In the higher mass stars, however, some objects have quite low 
$^{16}$O/$^{17}$O ratios and cannot be  consistent with the 
predicted trend. 

\begin{figure}[ht]
\begin{center}
\hspace{-0mm}
\includegraphics[height=45mm]{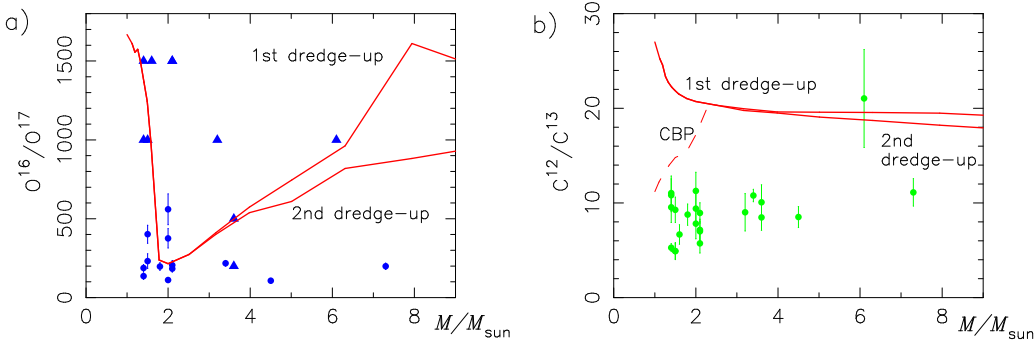}
\caption{ a) Observed  and predicted  
$^{16}$O/$^{17}$O ratios. b) The same for
$^{12}$C/$^{13}$C ratios }
\label{Fig3.eps}
\end{center}
\end{figure}

{\bf $^{12}$C/$^{13}$C Ratio:}
We confirm in Fig.\,3b that $^{12}$C/$^{13}$C ratios are mostly 
around 10.  The contradiction with the evolutionary models, which predict 
$^{12}$C/$^{13}$C\,$\approx 20$ (solid lines), has been a long-standing
puzzle. This dilemma may be resolved by assuming
a  deep circulation  below the bottom of the
convective zone, referred to as  {\it cool bottom processing} 
(\cite[Boothroyd \& Sackmann 1999]{Boothroyd99}), and its
prediction is also plotted in Fig.\,3b (dashed line). 
Our result, however, suggests that the extra mixing 
may not resolve yet the $^{12}$C/$^{13}$C puzzle,
even if  we make allowance for uncertainties in stellar masses. 
 
{\bf Concluding Remarks:}
We also planned to redetermine C abundances, 
 which showed somewhat different results
from the CO 1st (\cite[Tsuji 1986]{Tsuji86}) and 2nd  
(\cite[Tsuji 1991]{Tsuji91}) overtone bands,
 and details on this issue will be discussed  elsewhere.
Despite such a problem, the isotopic ratios can be determined more 
accurately than the elemental abundances. 
For this reason, isotopic ratios can be useful probes of 
evolutionary models and mixing processes in cool evolved stars.
The observed and predicted isotopic ratios, however, are still by no means  
consistent (Fig.\,3). Introduction of an extra-mixing may be needed,
 but it is vital to pursue a better theory of convection that may 
comprehend more efficient mixing.

\end{document}